# THERMODYNAMICS OF THE TRANSFORMATION OF GRAVITATIONAL WAVES INTO MATTER QUANTUMS FOR A VACUUM SPACE MODEL

J. A. Montemayor-Aldrete[1], M. López de Haro[2], J. R. Morones-Ibarra[3], A. Morales-Mori[4], Mendoza-Allende[1], E. Cabrera-Bravo[1] and A. Montemayor-Varela[5].

1. Instituto de Física, Universidad Nacional Autónoma de México, Apartado Postal 20-364, 01000 México, D. F.

2. Centro de Investigación en Energía, Universidad Nacional Autónoma de México, Apartado Postal 34, 06258 Temixco, Mor., México.

3. Facultad de Ciencias Físico-Matemáticas, Posgrado, Universidad Autónoma de Nuevo León, Apartado Postal 101-F, 66450 San Nicolás de los Garza, Nuevo León.

4. Centro de Ciencias Físicas, Universidad Nacional Autónoma de México, Apartado Postal 139-B, 62191 Cuernavaca, Morelos. México.

5. Centro de Mantenimiento, Diagnóstico y Operación Iberdrola, Polígono Industrial, El Serrallo, 12100, Castellón de la Plana, España.

**ABSTRACT**

It is shown that the entropy of low density monochromatic gravitational waves, waves required for the stabilization of the crystalline structure of vacuum cosmic space, varies with the volume in the same manner as the entropy of an ideal gas formed by particles. This implies that close enough to the big-bang event the energy of all the $10^{120}$ gravitational waves, under an adiabatic compression process, which stabilizes the crystalline structure of vacuum space behaves thermodynamically as though it is consisted of a number $n_B = 10^{80}$ of independent energy or matter quanta (neutrons).

# I    INTRODUCCION

Recently, several implications of a crystalline model for the vacuum cosmic space with lattice parameter of the order of the neutron radius have been analyzed [1, 2]. In the first paper [1] a formal analogy between dislocation creep and Relativistic Cosmology has been explored. One of the new equations obtained for cosmology has two characteristic distances: the present Universe radius, $R_{OU}$, and the lattice parameter which is roughly the neutron radius size $r_N$. In the second paper [2] Heisenberg's uncertainty principle was used to find the gravitational stability condition of the crystalline structure of vacuum space. This stability condition implies an equation which relates the temperature, $T(r)$, the radial distance, $r$, and the energy density, $U(r)$, of the gravitational waves under an adiabatic compression process due to their own interaction. Such equation is formally equal to one previously obtained by Gamow and coworker [3,4], once pair production has ceased, to predict the present microwave background temperature from the matter density. Gamow's equation arises from general relativity theory as applied to the big bang event.

In mathematical language, on the one hand, for the gravitational waves which stabilize the crystalline vacuum space structure under an adiabatic compression process due to their own gravitational interaction, we have:

$$\frac{T(r)}{T_{OU}} = \frac{R_{OU}}{r} = \left(\frac{U(r)}{U_{OU}}\right)^{1/3} , \qquad (1)$$

where $T_{OU} \sim 10^{27} K$ is the absolute temperature stabilization of gravitational waves at radius $R_{OU}$ and $U_{OU}$ is energy density of such waves at the same radius.

On the other hand, for the relativistic case we have that once pair production has ceased, $\rho$ the matter density varies simply as [3-5].

$$\frac{T_1}{T_o} = \frac{L_o}{L_1} = \left(\frac{\rho_1}{\rho_o}\right)^{1/3}, \qquad (2)$$

where $T_1$ and $T_o$ are absolute temperatures, $L_1$ and $L_o$ are radial distances. According to Penzias [5]: "If we take $T_1$ and $\rho_1$ to be the radiation temperature and matter density at the time of deuterium formation ($10^9$K and $10^{+5}$g/cm$^3$), we have the relation first used by Gamow to predict the present temperature of the microwave background from the matter density". Also Eqs. (1) and (2) are also formally similar with previous conclusions obtained by Homer Lane in (1869) called to Lane's theorem (as cited by Chandrasekhar [6]) for the uniform expansion of an ideal gas sphere, where particles of such gas are interacting gravitationally.

This equality between Eqs. (1) and (2) is meaningful because Eq. (1) is referring to gravitational waves for the crystalline vacuum cosmic space model and Eq. (2) refers to matter packages arising from the big bang. And by considering that both processes are adiabatic in nature (compression for the first one and expansion for the big bang) then it is possible, by energy conservation, to consider such event as a totally inelastic interaction or collision between incoming gravitational waves and an out coming stream of matter packages from the big bang. In other words, it is possible to imagine such process as a full transformation of such gravitational waves, which stabilize the crystalline structure of the vacuum cosmic space, into matter quanta during the first instants of the big bang.

This last hypothesis deserves further exploration because according to usual considerations (see Penrose [7]) the baryons number in the present Universe volume $V_{OU}$ is $10^{80}$; and the total energy inside $V_{OU}$ due to gravitational waves for the crystalline structure stabilization of the vacuum cosmic space is $E_{OU} = 10^{120} h \nu_{OU} = 10^{120} \frac{\eta c}{R_{OU}} = 10^{80} u_N$, where $u_N$, is the self energy for a neutron [2] and $c$ is the light speed.

This transformation of gravitational stabilizing waves of the vacuum space with crystalline structure into matter quanta (neutrons) could be studied in two complementary schemes. First a thermodynamic one, second a quantum scheme. This last scheme will be worked out in a separate paper.

In this paper the thermodynamic analysis will be made in two possible ways. First we study the entropy properties of the gravitational self-attracting waves, later an analysis on the Gibbs free energy changes, in the crystalline vacuum cosmic space, due to the transformation process of gravitational waves (before mentioned) into matter quanta will be made.

## 2. THEORY

Before the classical work due to Einstein concerning an heuristic point of view toward the emission and light transformation [8], theoretical physicists consider that a profound formal distinction exists between the theoretical concepts regarding gases and other ponderable bodies and the Maxwellian theory of electromagnetic process in the so - called empty space [8, 9].

Einstein by obtaining the asymptotic form for the monochromatic electromagnetic radiation entropy at low radiation density and large values of $v/T$; shows that such entropy varies with the volume in the same manner as the ideal gas entropy or a dilute solution. And he also shows that low density monochromatic radiation (within the validity range of Wien's radiation formula) behaves thermodynamically as though it consisted of a number of independent energy quanta or discrete energy corpuscles $hv$.

It is clear that at the present day is not possible to make experiments concerning gravitational waves, however there are many experimental data for electromagnetic waves; and because for weak gravitational fields, which corresponds to the linear region of the Einstein's field equations there is a strong analogy between Maxwell's and Einstein's equations one would expect that electromagnetic and gravitational waves have a similar behavior. Based in such consideration on the one hand, it is possible to see that the gravitational waves which stabilize the crystalline structure of the vacuum cosmic space obey the following behavior

$$s(v_g, T) = -\frac{U_V(v_g, T)}{\beta v_g}\left[\lambda n\left(\frac{U_V(v_g, T)}{\alpha v_g^3}\right) - 1\right] \quad (3)$$

where $s(v_g, T)$ is the entropy volumetric density due to gravitational waves of frequency $v_g$ in the limit of high ($v_g/T$) ratios. Here $U_V(v_g, T)$ is the gravitational waves energy density with frequency $v_g$ and temperature $T$. Also $\alpha$ and $\beta$ have the meaning ascribed to them by Einstein [8, 9].

In our case due to the adiabatic compression process occurring between the gravitational waves which stabilize the vacuum cosmic space of volume $V_{OU}$, energy is conserved and

$$E_{OU} = U_{OU} V_{OU} = E(v_g, T) = U_V(v_g, T) V(v_g, T) \quad (4)$$

And, also it is possible to define:

$$S_{OU}(v_g, T_{OU}) \equiv s(v_g, T_{OU}) V_{OU} \quad (5.a)$$

and

$$S(v_g, T) \equiv s(v_g, T) V(v_g, T) \quad (5.b)$$

Therefore by using Eq. (3) together with Eqs. (4) and (5) it is easy to show that

$$\Delta S_{gw} \equiv S(v_g, T) - S_{OU}(v_g, T_{OU}) = \frac{E(v_g)}{\beta v_g} \ln\left(\frac{V}{V_{OU}}\right) \quad (6)$$

On the other hand, when inside a volume $V$ we have a gas formed by an arbitrary number, $n_B$, of matter particles, considered in a first approximation without extension, which are so diluted that do not interact between them; then the entropy of such gas of particles can be described by,

$$\Delta S_{gas} = \frac{R}{N_a} \ln\left(\frac{V}{V_{OU}}\right)^{n_B} \quad (7)$$

$$= \frac{R n_B}{N_a} \ln\left(\frac{V}{V_{OU}}\right) \quad (8)$$

where $V < V_{OU}$. By comparing Eq. (6) with Eq. (8) it is clear that the entropy of monochromatic gravitational waves of sufficiently low density varies with the volume in the same manner as the ideal gas entropy or a dilute solution.

The compatibility condition between Eqs. (8) and (6) is:

$$E(v_g) = n_B h v_g \quad (9)$$

where $\beta$ has been taken as $(h/k)$, with $k$ as the Boltzmann's constant [8, 9]. The compatibility condition in a physical sense only admits the solution implied by the big bang where $n_B$ denotes the baryons number ($10^{80}$) inside $V_{OU}$, with a value for $v_g$ equal to $v_N$ (neutron frequency). And by taking into account Eq. (4), $E(v_g) = E_{OU}$. Then

$$E_{OU} = n_B h v_N = n_B u_N \quad (10)$$

In other words,

$$E_{OU} = 10^{120} \, h\nu_{OU} = 10^{80} \, u_N \qquad (11)$$

These equations have the following physical meaning. At the big bang, the energy of all the gravitational waves which stabilize the crystalline vacuum cosmic space against gravitational fluctuations behaves thermodynamically as though it consisted of a number $n_B$ of independent energy or matter quanta of magnitude $h\nu_N = u_N$.

But Eq. (11) allows another complementary interpretation:

$$u_N = 10^{40} \, h\nu_{OU} \qquad (12)$$

which means that the energy of $10^{40}$ gravitational waves for the stabilization of the crystalline structure of the vacuum cosmic space transforms into a matter quanta called neutron. This situation and the occurrence of a previous adiabatic compression process of gravitational waves, open up the possibility that elementary matter quanta could be conceptualized as a new kind of black-hole formed by gravitational waves circling around a center in a similar way to the De Broglie model of matter packages. This possibility will be examined elsewhere.

Equation (10) could be written in terms of the absolute temperature associated to such energy [8,9], namely

$$E_{OU} = n_B \, kT_{NF}, \qquad (13)$$

where $u_N$ has been rewritten as $kT_{NF}$, with $T_{NF}$ as the temperature for the formation of neutron anti-neutron pairs. $T_{NF} = 10^{13} K$ [10,11].

9Then from Eq. (13)

$$kn_B = \frac{E_{OU}}{T_{NF}} \tag{14}$$

And, by using that $E_{OU} = k\,T_{OU}$ [2] and by using Eq. (11)

$$k10^{120} = \frac{E_{OU}}{T_{OU}} \tag{15}$$

Equations (14) and (15) have the classical form of entropy function. By denoting $S_{OU}$ for $k10^{120}$ and $S_{NF}$ for $kn_B = k10^{80}$, we have $S_{OU} = E_{OU}/T_{OU}$ and $S_{NF} = E_{OU}/T_{NF}$, or

$$S_{OU} = 10^{40}\, S_{NF} \tag{16}$$

This result explains the low value of the starting Universe entropy at the big bang as requested by Penrose [6]. (The associated problem of bounds on the entropy of any object or system of maximum radius $R$ and total energy $E$ [12] as implied by our analysis will be studied in a future paper.

From information provided in this paper it is clear that the internal energy, $E_{OU}$, due to quantum fluctuations to stabilize the crystalline structure of vacuum cosmic space inside a volume $V_{OU}$ is $10^{-40}$ times the internal energy of the perfect crystalline structure of vacuum inside $V_{OU}$, $E_{vc}$, $\left(E_{OU} = 10^{-40}\, E_{vc}\right)$. In other words, in a thermodynamic scheme we are facing a fluctuation in internal energy of the crystalline structure of vacuum space.

In crystalline physics many of the processes occur under constant volume (here $\Delta V/V_{OU} \sim 10^{-40}$) and the Gibbs free energy per unit volume, $G_g$, is frequently simplified to read [13].

$$G_g = E - TS \tag{17}$$

where $E$ is internal energy, $T$ absolute temperature and $S$ entropy. With $E$ and $S$ as measured relatives to the perfect crystal. Because in our analysis we are facing a fluctuation in the internal energy of the crystalline structure of vacuum space in some conditions the change on $G_g$ will be greater than zero and in other situations the change on $G_g$ will be lower than zero, but in the average a value zero for the change on $G_g$ is expected.

Therefore in our case we have that for the starting of the gravitational stabilization process of the crystalline structure of vacuum $(G_g)_{OU}$ is:

$$(G_g)_{OU} = E_{OU} - T_{OU} S_{OU} \tag{18}$$

Where as before $E_{OU} = 10^{120} h\nu_{OU} = 10^{120} \dfrac{\eta c}{R_{OU}}, T_{OU} \approx 10^{-27} K, and\ S_{OU} = 10^{120} k$. Then

$$(G_g)_{OU} = E_{OU} - k10^{-27}10^{120} = 0 \tag{19}$$

And also for the physical situation where gravitational waves for gravitational stabilization of the crystalline structure of vacuum cosmic space transforms into matter quanta (specifically in neutrons). This means that $(G_g)_{NF}$ appears as

$$(G_g)_{NF} = E_{NF} - T_{NF} S_{NF} \tag{20}$$

where $E_{NF} = E_{OU}$ because of the adiabatic nature of the self-compression occurring between the previously mentioned gravitational waves, $T_{NF}$ is equal to $10^{13}K$ as before, and $S_{NF} = 10^{80}K$. With these values $(Gg)_{NF}$ becomes

$$(Gg)_{NF} = E_{OU} - k10^{13}\ 10^{80} \tag{21}$$

$$= 0 \tag{22}$$

Therefore $(Gg)_{OU} = (Gg)_{NF}$ means that the Gibbs' free energy of the crystalline structure of vacuum space due to quantum fluctuations for gravitational stabilization of vacuum space is equal to the Gibb's free energy of the crystalline structure when $10^{80}$ baryons (neutrons) are formed before the big bang begins.

3. **DISCUSSION AND CONCLUSSIONS**

3.1) It has been shown that the entropy of low density monochromatic gravitational waves, for the stabilization of the crystalline structure of vacuum space, varies with the volume in the same manner as the entropy of an ideal gas formed by particles. In other words, it has been shown that close enough to the big bang event the energy of all the $10^{120}$ gravitational waves (of frequency $\nu_{OU}$) which stabilizes the crystalline vacuum space structure behaves thermodynamically as though it consisted of a number $n_B = 10^{80}$ independent energy or matter quanta of magnitude each (neutrons) $u_N = h\nu_N = 10^{40} h\nu_{OU}$.

3.2) Equations (14) and (15) not only allow us to explain the low value of the starting of the Universe entropy (as requested by Penrose [7]); but also it has a possible connection with expressions for the entropy of black holes. Both equations resemble an expression for the entropy of a black hole, $S_{BH}$, due to Bekenstein [14] and Hawking [15] which if it is taken seriously [16] allows to exhibit that the first law of the thermodynamics dictates that black holes must have a temperature $T$, as follows

$$d(M_{BH} c^2) = T\, dS_{BH} \qquad (23)$$

where $M_{BH}$ is the total mass of the black hole. This equation could be expressed as

$$dS_{BH} = \frac{d(M_{BH}c^2)}{T} \tag{24}$$

It is clear that Eqs. (14) and (15) used together with Eq. (11) have the same form and physical meaning of Eq. (18) and therefore the possibility exists that elementary particles could be conceptualized as a new kind of black hole. This possibility will be studied in a following paper.

3.3) Relative to Eqs. (20) and (21) in our scheme neutrons and antineutrons are formed at the same rate. However because of the gravitational stress gradient pattern existing around the source region, all the matter travels in one direction and all the antimatter travels in the opposite direction so there are no annihilation events during this stage. Similar considerations about matter and antimatter productions processes have been used before by other authors (see Hawking [17] where he proposes a mechanism for black-hole evaporation). In our scheme, massive annihilation events between matter and antimatter will occur at the last stage of the expansion cycle of matter on the present fluctuation of the internal energy of the crystalline structure of vacuum space. This point will be fully explained in a further paper.

3.4) In nature crystals are never perfect, and it is well understood that they may contain many defects. One of the most important is called a vacancy. When vacancies are introduced into the crystalline structure the internal energy of the crystal increases and the entropy of the crystal increases too, but the Gibbs free energy diminishes until an equilibrium value is reached.

In our case the situation is different because of the role that gravitation forces play in relation to the crystalline structure of the vacuum cosmic space. Equations (19) and (21) and the results obtained in reference [2], allow us to consider, according to Boltzmann's analysis [18], that the degraded energy of the crystalline vacuum cosmic space of volume $V_{OU}$, which is in the most probable form of gravitational waves with frequency $\nu_{OU}$, lead to mutual gravitational interactions giving place to a "visible" and coordinate motion which behaves like heat of very high temperature that can be completely transformed into work. Not only $E_{OU} = E_{NF}$ but all the energy $E_{OU}$ of gravitational compressed waves transforms into matter quanta with total energy $E_{NF} = 10^{80} U_n$. This situation is similar to the one which has been observed in plastic deformation [19] where coherent elastic waves generated during an unloading process, travel along a deformed crystal to concentrate at dislocation sources giving place to a full transformation of the energy of the elastic waves into crystal defects called dislocations. In other word, during the plastic deformation process which occurs during an unloading test, energy in the form of coherent elastic sound waves, which have a very high entropy, give place to the production of dislocations (which have a very low configuration entropy) but have the same stored elastic energy that the original elastic waves.

Additionally, if we take into account that gravitational long range cosmological stresses and quantum effective forces acting on the crystalline structure of vacuum do not appear in an explicit way in Eqs. (19) and (21) it is possible to arrive at an apparent paradoxical situation where Eqs. (19) and (21) refer not to an absolute minimum in the Gibbs' free energy but to a relative minimum, due to the following reasons. On the one hand, as already mentioned, in the case of Eq. (19 the gravitational attraction between the stabilization gravitational waves provides a mechanism for the increases in the Gibbs' free energy of the system and a decrease occur when such waves transforms into baryons the Gibbs' free energy attaining its initial value, Eq. (21). After such process occurs, the dispersion of baryons through crystalline vacuum space begins and the Gibbs' free energy diminishes with the expansion until matter and antimatter (which were produced in the neighboring Universe cell and travel in opposite sense to matter facing the same cosmic gravitational stress gradient) "collides" and annihilate liberating the gravitational stored energy and leading to the rise of free energy to the level of $E_{OU}$.

On the other hand, from the physical situation existing previously to the fulfillment of Eq. (21) we have a transitory "fire ball" which decreases in radial distance and increases its temperature as the last stage of the adiabatic compression process proceeds on the gravitational waves which stabilize the crystalline structure of vacuum space. Such situation is characterized by a very high grade of instability. The gravitational energy of such waves compressed in the very last stage of the adiabatic process before mentioned causes a very high compressed matter quantum which produces huge expansion forces due to the uncertainty principle, giving place to a very fast inflation of the structural components of the very hot gases out of thermal equilibrium. (The factor of such sudden inflation, which will be explained later, is $10^{40}$). This physical situation correspond to a minimum entropy situation; see Eq. (16), which states that the starting entropy is about $10^{-40}$ of the final entropy of the Universe. We consider that the situation here described is in agreement with Penrose's analysis about the expected thermodynamical conditions to be fulfilled at the beginning of the expansion cycle [7].

## ACKNOWLEDGEMENTS

We acknowledge to the librarian Technician G. Moreno for her stupendous work and to O. N. Rodríguez Peña for her patient and skilful typing work.